\documentclass[aps,prl,preprint,superscriptaddress]{iopart}

\usepackage{latexsym}
\usepackage{amssymb}
\usepackage{graphicx}  %  include figure files
\usepackage{epsfig}    %  files .eps
\usepackage{dcolumn}   %  align tble columns on decimal points
\usepackage{bm}        %  bold math

\usepackage{color}  

\newcommand{\Epsilon}{\mathcal{E}}

\begin{document}

% TITLE, AUTHORS AND ABSTRACT
%%%%%%%%%%%%%%%%%%%%%%%%%%%%%%%%%%%%%%%%%%%%%%%%%%%%%%%%%%%%%%%%%%%%%%%%%%%%%%%%%%%%%%%%%%%%%%%%%%%
\title[Ion beams by irradiating a nano-structured target with a PW laser pulse]{High-quality ion beams by irradiating a nano-structured target with a petawatt laser pulse}

\author{M. Grech}
\address{Max-Planck-Institute for the Physics of Complex Systems, D-01187 Dresden, Germany}

\author{S. Skupin}
\address{Max-Planck-Institute for the Physics of Complex Systems, D-01187 Dresden, Germany}
\address{Institute of Condensed Matter Theory and Solid State Optics, Friedrich-Schiller-University Jena, D-07743 Jena, Germany}

\author{R. Nuter, L. Gremillet and E. Lefebvre}
\address{CEA, DAM, DIF, F-91297 Arpajon, France}

\date{\today} 

\ead{mickael.grech@gmail.com}

\begin{abstract}

We present a novel laser-based ion acceleration scheme, where a petawatt circularly polarized laser pulse is shot on an ultra-thin (nano-scale) double-layer target. Our scheme allows the production of high-quality light ion beams with both energy and angular dispersion controllable by the target properties. We show that extraction of all electrons from the target by radiation pressure can lead to a very effective two-step acceleration process for light ions if the target is correctly designed. Relativistic protons are predicted with pulse powers of a few petawatt. Careful analytical modeling yields estimates for characteristic beam parameters and requirements on the laser pulse quality, in excellent agreement with one and two-dimensional Particle-in-Cell simulations.

\end{abstract}

%\pacs{?}
\maketitle
%%%%%%%%%%%%%%%%%%%%%%%%%%%%%%%%%%%%%%%%%%%%%%%%%%%%%%%%%%%%%%%%%%%%%%%%%%%%%%%%%%%%%%%%%%%%%%%%%%%

%%%%%%%%%%%%%%%%
% INTRODUCTION %
%%%%%%%%%%%%%%%%
\section{Introduction}\label{sec1}

Generation of high-energy ion beams by interaction of an ultra intense laser pulse with a solid target is one of today's hot topics in laser-plasma interaction. Such ion beams have properties making them very interesting for a wide range of applications, such as proton radiography~\cite{borghesi_POP_02}, fast ignition in the context of inertial fusion~\cite{roth_PRL_01,temporal_POP_02,naumova_PRL_08}, or hadron-therapy~\cite{koshorovEJP_bulanovPPR}. While generation of ions with energies up to several tens of~MeV has already been demonstrated~\cite{snavely00_robson06}, controlling their energy distribution remains a crucial issue for most applications. 

Different mechanisms of ion acceleration have been proposed depending on whether ions originate from the front-side (irradiated by the laser) or the rear-side of the target. Ion acceleration at the target front-side occurs mainly in the electric field resulting from electron sweeping at the front of the laser pulse~\cite{sentoku_POP_03}, leading to creation of collisionless electrostatic shocks~\cite{shock_acceleration,macchi_PRL_05} or solitary waves~\cite{solitary_wave}. At the target rear-side, ion acceleration occurs in the strong electrostatic field resulting from charge separation due to hot electrons escaping into vacuum. This mechanism, referred to as target normal sheath acceleration (TNSA)~\cite{wilks_POP_01}, is the dominant process of ion acceleration for currently available (moderately relativistic) laser intensities~\cite{fuchs_PRL_05}. TNSA provides ion beams with interesting properties, such as good laminarity, small aperture angle (a few degree) and relatively large efficiency of energy conversion (a few percents) from the laser pulse to the ions. However, the resulting ion beams have a characteristic broad (quasi-thermal) spectrum with a sharp cut-off at maximal energy. 

Several proposals have been made how to control the energy distribution of laser-created ion beams. Recently, efficient acceleration with relatively small energy dispersion has been observed in numerical simulations where a circularly polarized (CP) laser pulse was focused on a thin target~\cite{zhang_POP_07,klimo_PRSTAB_08,robinson_qiao,yan_rykovanov_eliasson,macchi_PRL_09}. Using CP light indeed allows to strongly reduce electron heating~\cite{macchi_PRL_05} and therefore prevents TNSA and the associated broadening of the ion spectrum. Ion acceleration then follows from a front-side mechanism referred to as laser-piston or light-sail acceleration. The whole target is accelerated as a neutral bunch, resulting in a quasi-monochromatic ion energy distribution. However, due to the non-homogeneous field in the piston, the presence of low-energy ions in the beam cannot be avoided. 

Moreover, alternative methods based on multi-species (homogeneous or multilayered) targets have been proposed to control the ion spectrum~\cite{esirkepov_PRL_02,schwoerer_NATURE_06,hegelich_NATURE_06,albright_PRL_06,teravetisyan_PRL_06,brantov_POP_06}. In particular, double layer targets have attracted a lot of interest as they allow to increase ion beams monochromaticity~\cite{schwoerer_NATURE_06,hegelich_NATURE_06}. In the proposed schemes, a laser pulse is focused on the first target layer that consists of heavy, highly charged, ions. Electrons gain energy in the laser field and are either heated or completely extracted from the laser focal spot. In the first case, ions of the second layer are accelerated in the ambipolar electrostatic field created at the rear-side of the first ion layer by hot electrons~\cite{albright_PRL_06}. The possibility to generate 1.3~MeV proton beams with energy dispersion $\sim 25\,\%$ and 3~MeV~carbon beams with energy dispersion $\sim 17\,\%$ using double layer target has already been demonstrated experimentally by Schwoerer~{\it et al.}~\cite{schwoerer_NATURE_06} and Hegelich~{\it et al.}~\cite{hegelich_NATURE_06}, respectively. The mechanisms behind these observations are similar to TNSA, and generation of hot electrons is a dominant process under present experimental conditions. This rather complex mechanism makes the control of the ion beam properties non trivial. In the second case, where electrons are removed from the target, light ions are accelerated in the strong electrostatic field of the expanding first ion layer. This is the so-called regime of directed Coulomb explosion (DCE)~\cite{esirkepov_PRL_02,DCE}. While ion acceleration in the ambipolar field is highly sensitive to the hot electron temperature and thus to the laser parameters, ion acceleration by DCE depends only on the target properties, thus allowing a better control of the ion source. Nevertheless, no efficient mechanism has yet been proposed to expel all electrons from the target and achieve optimal DCE.

In this paper, we present an elaborated laser-based ion acceleration scheme which allows to control energy dispersion, potentially below 10~$\%$. This scheme relies on the complete, laser-induced, removal of electrons from an ultra-thin double layer target. To make electron sweeping by the laser radiation pressure more effective, we make use of a CP laser beam. One-dimensional (1D) and two-dimensional (2D) numerical simulations are performed with the particle-in-cell (PIC) code CALDER~\cite{lefebvre_NF_03}. They show that complete extraction of electrons from the target is possible above a threshold intensity that depends mainly on the target areal charge. The resulting ion acceleration from the double layer target is then discussed using both numerical simulations and analytical modeling. First, light ions making the second target layer are accelerated in the quasi-homogeneous electrostatic field created between the first ion layer (that consists of heavy, highly charged, ions) and the forward-going electron cloud. This first stage, referred to as linear plasma acceleration (linPA), ends when electrons are pushed far enough from the heavy ion layer. The subsequent acceleration phase of light ions depends on whether they have acquired or not relativistic velocities in the linPA stage. Especially, it is shown that low-energy ions can gain further energy in a stage similar to DCE, while high-energy, relativistic, ions gain most of their energy in the linPA stage of acceleration. The final energy and energy dispersion are shown to depend mainly on the target composition and geometry. 

Because this novel mechanism of light ion acceleration depends mainly on the target properties, it is thus much less sensitive to the inevitable shot-to-shot fluctuations of the laser parameters than TNSA-based schemes. The ability to control the ion beam properties by a careful design of the target makes this acceleration process particularly interesting for applications requiring high-quality ion beams, such as hadron-therapy.

%%%%%%%%%%%%%%%%%%%%%%%%%%%%%%%%%%%%%%%%%%%%%%%%%%%%
%    THE linPA: THEORY                             %
%%%%%%%%%%%%%%%%%%%%%%%%%%%%%%%%%%%%%%%%%%%%%%%%%%%%
\section{The linear plasma accelerator (linPA): Theory}\label{sec2}

\begin{figure}
\begin{center} \includegraphics[width=6cm]{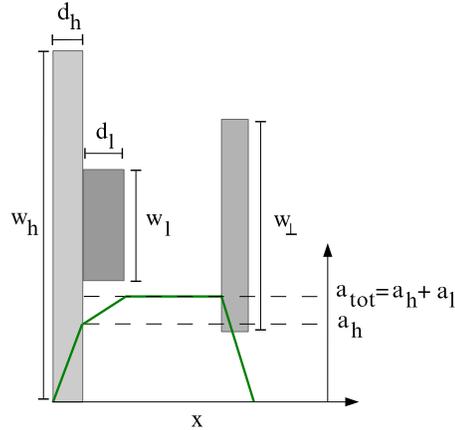}\end{center}
\caption{Characteristic double-layer target used in this paper and spatial distribution of the electrostatic field along the $x$-direction (green curve). Quantities $w_h$, $d_h$ and $w_l$, $d_l$ denote the transverse size and thickness of the heavy and light ion layers, respectively. The electron bunch (on the right) has the transverse size $w_{\perp} \sim {\rm min}\{w_h,w_L\}$, where $w_L$ is the width of the laser focal spot. The laser pulse propagates from left to right.}
\label{fig0}
\end{figure}

Our characteristic target (see Fig.~\ref{fig0}) consists of a first layer of heavy, highly charged ions with mass $m_l \gg m_p$ ($m_p=1836$ is the proton mass) and charge $Z_h \gg 1$, atomic density $n_h$ and thickness $d_h$. In this paper, mass, densities, charges and distances are normalized to the electron mass $m_e$, critical density $n_c = \epsilon_0\,m_e\,\omega_L^2/e^2$ ($\epsilon_0$ is the permittivity of vacuum, $e$ is the electron charge and $\omega_L$ is the laser frequency), electron charge $e$ and inverse laser wave number $k_L^{-1}=c/\omega_L$, respectively. A second layer, with density $n_l$ and thickness $d_l$, contains the light ions whose acceleration is considered. As will be discussed later in the paper, these ions must have a charge-over-mass ratio larger than the one of the so-called heavy ions. This is usually the case for species with $m_l \ll m_h$, first because of the neutron contribution to the mass of the nucleus, and second because the heavy ions may not be fully ionized. This target is irradiated by a CP laser beam at relativistic intensity. In what follows, the laser field amplitude\footnote{Normalization are chosen so that $I_L\,\lambda_L^2=1.38\,a_L^2 \times 10^{18}\,{\rm W/cm^2\,\mu m^2}$, with $I_L$ the laser intensity and $\lambda_L$ the laser wavelength.} $a_L = E_L/E_C > 1$, as well as all other electric fields, are given in units of the Compton field $E_C = m_e\,\omega_L\,c/e$.

\subsection{Electron response to laser radiation pressure: creation of the linPA}

In a first stage, electrons are quickly pushed forward into the target by radiation pressure $\Pi_L = (1+R-T)\,a_L^2/2$, where $\Pi_L$ is given in units of $n_e\,m_e\,c^2$, and $R$ and $T$ are the laser reflectivity and transmittance. As electrons penetrate deeper into the target, an electrostatic field is built up at the target front side. If the target is thick enough, a quasi-equilibrium between electrostatic and radiation pressures can set in. This process has been observed in different studies related to hole-boring~\cite{shock_acceleration,macchi_PRL_05} or light-sail acceleration~\cite{zhang_POP_07,klimo_PRSTAB_08,robinson_qiao,yan_rykovanov_eliasson,macchi_PRL_09,esirkepov_PRL_04}, using both linearly polarized (LP) or CP light.  

In this paper, we consider the case where radiation pressure is too high to be balanced by the maximum electrostatic pressure $a_{tot}^2/2$ (where $a_{tot}=a_h+a_l$ is the maximum electrostatic field due to complete electron-ion separation; $a_h=Z_h\,n_h\,d_h$ and $a_l=Z_l\,n_l\,d_l$ are the electrostatic fields due to the bare ions of the first and second layers, respectively). When the laser field amplitude $a_L$ becomes larger than $\alpha\,a_{tot}$, with $\alpha$ of the order of unity\footnote{The parameter $\alpha$ on laser intensity accounts for relativistic corrections, spatial and temporal profile of the laser pulse, etc.}, electrons are completely extracted from the target\footnote{Let us note that, considering a plasma with ion density $\ll m/Z^2$ and a laser field amplitude $\ll m$ (where $m$ and $Z$ are the ion mass and charge, respectively), the ion layer is transparent to the laser beam.}. To our knowledge, this behavior was reported for the first time in Ref.~\cite{kulagin} where the possibility to generate dense electron bunches with ultra-intense laser pulses with sharp rising edge is discussed. More recently, Refs.~\cite{yan_rykovanov_eliasson,macchi_PRL_09} have reported similar electron behavior when investigating the transition from RPA to Coulomb explosion of a monolayer target. It is interesting to note that, for a given target electron density, complete extraction of all electrons indeed requires the target thickness to be smaller than the so-called optimal thickness for RPA.

For laser field amplitude $a_L \gg 1$, electrons become relativistic almost instantaneously and the separation time scale is $t_c \sim d_h + d_l$ (in units of $\omega_L^{-1}$). This time scale is much shorter than the pulse duration and all characteristic times of ion acceleration. 

\subsection{Ion acceleration in the linPA}

Right after the electrons are separated from the ions, the target exhibits a capacitor-like structure, with a uniform electrostatic field of amplitude $a_{tot}$ built up between the electron and ion layers. In this Section, we restrict our study to a 1D problem, thus the electrostatic field amplitude does not depend on the distance between the ion and electron layers. Limiting multi-dimensional effect are discussed later in Sec.\ref{sec3}. 

The electric field seen by the heavy ions of the first layer is not homogeneous. Assuming a flat-top ion density profile, it increases linearly from $0$ to $a_h$ so that the velocity of a given ion depends on its initial location, thus yielding a large energy spread. If the charge-over-mass ratio of the light ions is larger than the one of heavy ions, there are no intersections between the trajectories of heavy and light ions, and the electrostatic field in the second ion layer varies linearly from $a_h$ up to $a_{tot}$. With $a_l \ll a_h$, this field is quasi-homogeneous and small energy dispersion for light ions is expected. 

In 1D geometry and assuming complete electron expulsion, light ions see a constant, quasi-homogeneous, accelerating field. Relativistic equations of motion for the slowest and fastest light ions, experiencing the accelerating fields $a_h$ and $a_{tot}$ respectively, can thus be solved exactly. Extracting the ion mean energy $\Epsilon$ and energy dispersion $\Delta\Epsilon$ at time $t$ after creation of the linPA is straightforward:
\begin{eqnarray}
\label{eq1} \Epsilon       \sim m_l\,\left[ \sqrt{1+t^2/t_r^2} - 1\right]\,, \\
\label{eq2} \Delta\Epsilon \sim m_l\,\left[ \sqrt{1+\big(1+a_l/a_h\big)^2\,t^2/t_r^2} - \sqrt{1+t^2/t_r^2}\right]\,,
\end{eqnarray}
where energies are normalized to the electron rest energy $m_e\, c^2$ and times to $\omega_L^{-1}$. 

The characteristic time $t_r= m_l/(Z_l\,a_h)$ denotes the time required for light ions to gain relativistic energies. In the limit $t \ll t_r$, light ions have non-relativistic velocities and the mean energy and relative energy dispersion are $\Epsilon^{(c)} \sim m_l\,t^2/(2\,t_r^2)$ and $(\Delta\Epsilon/\Epsilon)^{(c)} \sim 2\,a_l/a_h$, respectively. On the contrary, if the linPA can be maintained over times $t \gg t_r$, light ions obtain ultra-relativistic velocities. Their mean energy then evolves as $\Epsilon^{(ur)}\sim m_l\,t/t_r$, while their relative energy dispersion is $(\Delta\Epsilon/\Epsilon)^{(ur)} \sim a_l/a_h$. In both limits, the relative energy dispersion remains small if the electrostatic field $a_l$ due to light ions is small compared to the accelerating field $a_h$ (due to heavy ions), and it can be controlled by adjusting the target properties. The acceleration scheme then suffers limitations similar to any linear accelerator: controlling the relative energy dispersion limits the areal density of accelerated ions $\sigma_{\perp} = n_l\,d_l$ (in units of $n_c/k_L$). Nevertheless, it is shown in what follows that satisfactory values of $\sigma_{\perp}$ can be obtained using laser field amplitude $a_L \sim 100$. 

It is interesting to note that the characteristic time $t_r$ as well depends only on the target parameters. In particular, for a given charge-over-mass ratio $Z_l/m_l$, $t_r$ depends only on the amplitude of the accelerating field $a_h$. Obviously, the larger $a_h$, the shorter $t_r$ and the higher the light ion energy at a given time $t$. Moreover, we want to emphasize here that the main limitation to the ion energy follows from multi-dimensional effects. While these effects are discussed in more detail in Sec.~\ref{sec3}, it is worth pointing out at this stage that they set in after electrons are pushed on a distance larger than either the laser focal spot diameter $w_L$ or the transverse width $w_h$ of the first target layer. Assuming relativistic electrons, this defines a time $t_{linPA} \lesssim w_{\perp}$ (where $w_{\perp} = {\rm min}\{w_L,w_h\}$) during which light ions can be efficiently accelerated in the electrostatic field $a_h$. The condition for obtaining relativistic ions on a time $t_{linPA}$ defines a minimum value for the accelerating field $a_h^{(r)} \sim m_l/(Z_l\,w_{\perp})$. Considering that the condition for charge separation ($a_L > \alpha\,a_h^{(r)}$) defines a threshold for the laser field, and noting that $P_L=a_L^2\,w_{\perp}^2$ defines the effective laser power on the target (in units of $I_C/k_L^2$), one can define a characteristic laser power $P_L^{(r)} = \alpha^2\,m_l^2/Z_l^2$ required to obtain relativistic light ions in the linPA. This characteristic power depends only on the light ions charge-over-mass ratio and it is typically of a few PW for protons.

%%%%%%%%%%%%%%%%%%%%%%%%%%%%%%%%%%%%%%%%%%%%%%%%%%%%
%    THE linPA: 1D NUMERICAL SIMULATIONS           %
%%%%%%%%%%%%%%%%%%%%%%%%%%%%%%%%%%%%%%%%%%%%%%%%%%%%
\section{The linPA: One-dimensional numerical simulations}\label{sec2bis}

\subsection{Numerical results vs. theoretical predictions}

One-dimensional simulations using the PIC code {\textsc CALDER} have been performed to explore this new regime of ion acceleration and to test theoretical predictions. In these calculations, a CP laser beam with field amplitude $a_L=100$ ($I_L \sim 1.4 \times 10^{22}\,{\rm W/cm^2}$ at a wavelength $\lambda_L = 1~{\rm \mu m}$) is focused at normal incidence on an ultra-thin (few tens of nm) double layer target. In a first attempt, a flat-top temporal laser intensity profile is chosen in order to simplify comparison to analytical estimates. 

The first target layer is made of carbon ($Z_h=6, m_h=12\,m_p$) with atomic density $n_h = 58$ ($\sim 6.4 \times 10^{22}\,{\rm cm^{-3}}$) and thickness ranging from $d_h=0.04$ up to $0.12$ ($6.4$ to $19.1$~nm). The second layer contains only hydrogen ($Z_l=1$, $m_l=m_p$) with density $n_l = 5.8$ ($\sim 6.4 \times 10^{21}\,{\rm cm^{-3}}$). Its thickness, $d_l = 0.04$ - $0.7$ (6.4 to 111~nm), is adjusted to control the energy dispersion of the resulting ion beam as well as its areal density $\sigma_{\perp}=n_l\,d_l$. The foil has step-like density profile in each layer. We assume complete ionization at the beginning of the simulation and the initial electron temperature is 1~keV.

In the following simulations, the numerical domain is $7\,\lambda_L$ long with mesh size $dx = 2.5 \times 10^{-3} \sim \lambda_{De}^{(h)}$, where $\lambda_{De}^{(h)}=\sqrt{T_e/(Z_h\,n_h)}$ is the normalized Debye length in the first target layer and $T_e \sim 1/511$ is the normalized initial electron temperature. The simulation duration is $5$ laser periods and the time step $dt \sim dx/2$. The laser propagates from the left to the right and reaches the target $\sim 1\,\tau_L$ after the beginning of the simulation ($\tau_L$ is the optical cycle). This time is referred to as the time zero. Entrant (absorbing) and absorbing (absorbing/reinjecting) boundary conditions are used for the electromagnetic field (particles) at the left and right edges of the simulation box, respectively. 

\begin{figure}
\begin{flushright} \includegraphics[width=14cm]{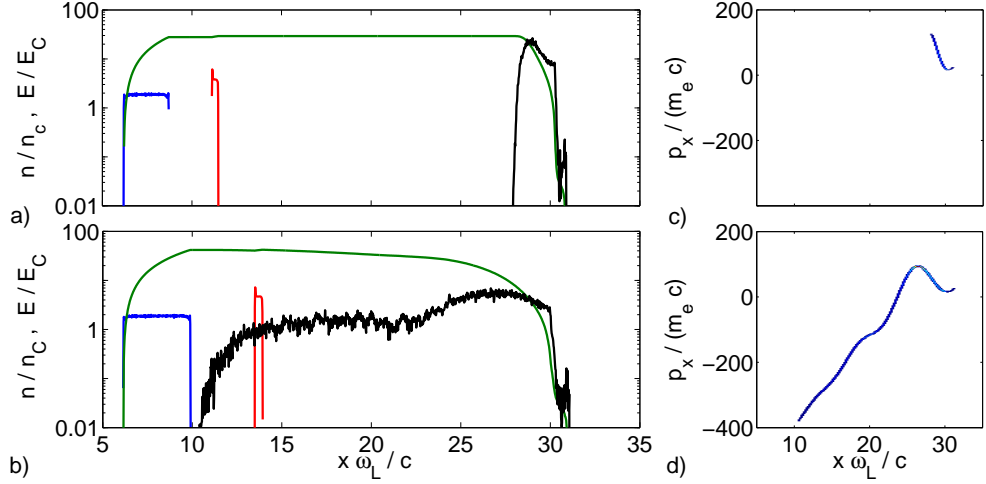}\end{flushright}
\caption{a)~Snapshot at $t=4\,\tau_L$ after the beginning of the interaction of the ion (blue: carbon, red: hydrogen) and electron (black) density and electrostatic field (green). The laser propagates from left to right. The target thickness is $d_h = 0.08$ ($\sim$12.7~nm). b)~Idem but with the target thickness $d_h = 0.12$ ($\sim$19.1~nm). c)~Electron momentum in the direction of the laser propagation versus position for parameters of panel a. d)~Idem but for parameters of panel b. }
\label{fig1}
\end{figure}

Figure~\ref{fig1} shows a snapshot of the ion and electron densities and the electrostatic field four laser periods after the beginning of the interaction, in the case where $d_l = 0.235$ ($\sim 37.4$ nm) and for two different thicknesses of the first layer $d_h = 0.08$ ($\sim$12.7 nm) and $d_h = 0.12$ ($\sim$19.1~nm), see Figs.~\ref{fig1}a~and~b, respectively. We can observe that, in the case of the thinnest target (Fig.~\ref{fig1}a), electrons are pushed as a compact bunch. As confirmed in the phase-space in Fig.~\ref{fig1}c, all electrons are pushed forward at relativistic velocities. For the thickest target, however (Figs.~\ref{fig1}b and~\ref{fig1}d), electrons are not pushed as a compact bunch anymore. Figure~\ref{fig1}d shows that, in this case, the electron dynamics is rather complex. Because the laser field does not strongly exceed the electrostatic field due to electron-ion separation, some electrons are pushed forward by the laser pulse while others are accelerated backward in the strong electrostatic field. A similar behavior has already been reported in the laser-piston regime by Naumova~{\it et al.}~\cite{naumova_PRL_08}. As demonstrated by these authors, the backward accelerated electrons may interact with the incident laser pulse, experience important radiative friction~\cite{steiger_PRA_72} and thereafter be slowed down. However, this effect is  not taken into account in our simulations. Later in this Section (see Sec.~\ref{temp}), more details are provided concerning the electron behavior and their effects on ion acceleration. 

Figures~\ref{fig1}a and~\ref{fig1}b also show that carbon ions are accelerated in an inhomogeneous (linearly varying) electrostatic field, and hence get smeared out along the laser propagation direction. On the contrary, protons see a quasi-homogeneous field, whose amplitude, $\sim 28$ ($\sim 90~$TV/m, see Fig.~\ref{fig1}a) and $\sim 42$ ($\sim 130~$TV/m, see Fig.~\ref{fig1}b), is in excellent agreement with the analytical prediction $a_{tot} \sim a_h=Z_h\,n_h\,d_h$. Moreover, because the charge-over-mass ratio is larger for protons than carbons, the proton layer is quickly separated from the carbon one. 

\begin{figure}
\begin{flushright} \includegraphics[width=14cm]{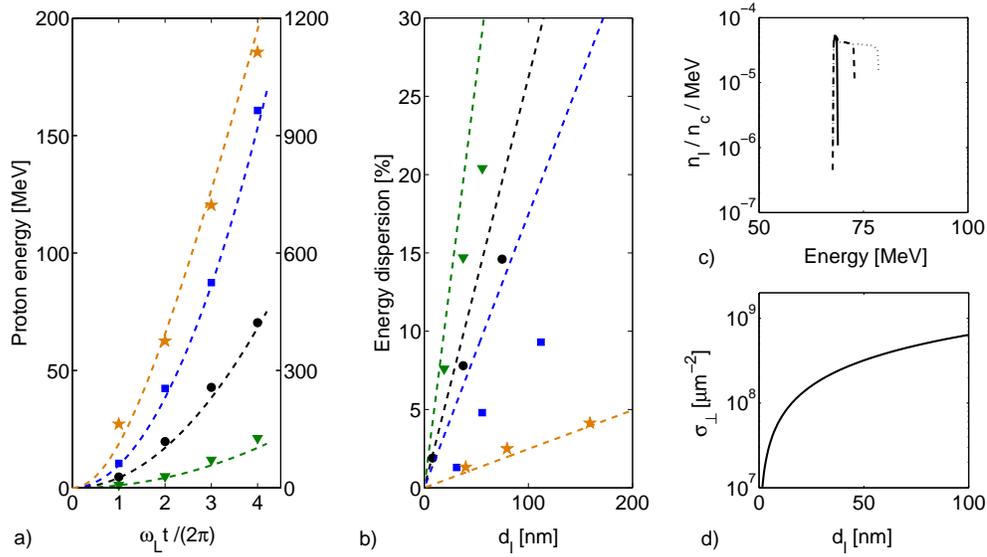} \end{flushright}
\caption{a)~Temporal evolution of the ion energy for different first layer thicknesses, $d_h = 0.04$ ($\sim$6.4~nm) (green, triangles), $d_h = 0.08$ ($\sim$12.7~nm) (black, circles) and $d_h = 0.12$ ($\sim$19.1~nm) (blue, square) and for parameters allowing relativistic ion generation (orange, star). The right vertical axis provides the energy corresponding to this last simulation. Dashed curves account for analytical estimates from Eq.~(\ref{eq1}). b)~Relative energy dispersion $\Delta\Epsilon / \Epsilon$ in $\%$ versus the second layer thickness $d_l$ for target parameter corresponding to panel a. c)~Energy spectrum of protons in the case where $d_h=0.08$ ($\sim$18.8~nm) and for three different second layer thicknesses: $d_l=0.05$ ($\sim$6.2~nm) (solid line), $d_l=0.25$ ($\sim$30~nm) (dashed line) and $d_l=0.5$ ($\sim$60~nm) (dotted line). d)~Dependence of the areal density of the proton beam on the second layer thickness $d_l$.}
\label{fig2}
\end{figure}

Figure~\ref{fig2} summarizes the proton beam properties obtained in numerical simulations and compares them to theoretical predictions. Temporal evolution of proton energy is shown in Fig.~\ref{fig2}a for different thicknesses of the first layer for parameters previously detailed. After a few laser cycles only, proton energies up to 150~MeV are obtained, in excellent agreement with analytical predictions from Eq.~(\ref{eq1}). Complementary simulations are also presented considering a carbon foil with thickness $d_h=0.27$ ($\sim 43$~nm) and solid density $n_h=91$ ($10^{23}\,{\rm cm^{-3}}$) irradiated by a CP laser with field amplitude $a_L=400$ (orange curves). Under such conditions, protons undergo a strong acceleration and can gain relativistic energies in only a few optical cycles. Here, protons with energy larger than 1~GeV are obtained, as expected from analytical modeling. This confirms that the linPA allows to generate high energy ion beams with high power laser on only a few laser cycles, that is without requiring exceedingly large laser energies. 

As previously underlined, the energy dispersion of the ion beam can be tuned by controlling the ratio $a_l/a_h$, i.e., the target properties. Figure~\ref{fig2}b presents the energy dispersion obtained in numerical simulations for different ratios $a_l/a_h$. As expected, the proton beam monochromaticity can be considerably improved by decreasing the thickness of the second layer, whereas the ion energy, which depends mainly on the first layer properties, is not modified. Energy dispersions of the order of $2~\%$ are obtained, which is very attractive for possible medical applications. Moreover, we point out that estimates from Eqs.~(\ref{eq1}) and (\ref{eq2}) compare well with numerical results. Actually, they may even overestimate them in those cases where some electrons partly neutralize the proton bunch, reducing the field inhomogeneity (Fig.~\ref{fig1}b). Last but not least, we want to emphasize that, in contrast to acceleration by radiation pressure (see e.g.~Refs.~\cite{klimo_PRSTAB_08,yan_rykovanov_eliasson,macchi_PRL_09}), and as can be observed in Fig.~\ref{fig2}c, all protons are contained within the monochromatic peak. 

The drawback of the low energy dispersion permitted by this method is the low areal number of accelerated protons. However, as shown in Fig.~\ref{fig2}d for typical parameters of this study, $\sigma_{\perp}=n_l\,d_l$ up to a few $10^9$ particles$/{\rm \mu m^2}$ can be reached, while keeping energy dispersion to a few percent only. 

To demonstrate acceleration of light ions other than protons, 1D simulations have been performed to investigate the generation of quasi-monochromatic carbon ion beams. Let us consider a $\lambda_L = 0.52\,{\rm \mu m}$-laser pulse with field amplitude $a_L=300$ ($I_L \sim 3.3\times 10^{22}\,{\rm W/cm^2}$) focused onto a 5~nm-thick ($d_h=0.06$) gold target at solid density $n_h=14.8$ ($\sim 5.9\times 10^{22}\,{\rm cm^{-3}}$). Because ionization is not accounted for in our code, we discuss two sets of simulations where gold is either completely ionized ($Z_h=79$ leading to $a_h \sim 70.2$) or electrons of the three inner shells remain bound ($Z_h=51$ leading to $a_h=45.3$). A fully ionized, thin, carbon layer with density $n_l=25$ ($\sim 10^{23}\,{\rm cm^{-3}}$) is placed at the rear side of the gold layer. Its thickness is adjusted so that relative energy dispersion of the carbon ion beam remains $\sim 10\,\%$: $d_l=0.045$ ($\sim 4~$nm) for the Au$^{79+}$ layer, and $d_l=0.030$ ($\sim 2.5~$nm) for the Au$^{51+}$ layer. As a result, the generation of carbon beams with relative energy dispersion of the order of $15\,\%$, and energy 190~MeV$/$nucleons (using Au$^{79+}$) and 80~MeV$/$nucleons (using Au$^{51+}$), is observed, in good agreement with theoretical predictions from the linPA model. This confirms that the proposed mechanism of acceleration can be applied to ions heavier than protons. However, it requires that the ions of the first target layer have a smaller charge-over-mass ratio than the ions whose acceleration is considered. Here, the gold ion charge-over-mass ratio is $\sim 0.40$ and $\sim 0.26$ for Au$^{79+}$ and Au$^{51+}$, respectively, and $0.5$ for the fully ionized carbons.

\subsection{Influence of the laser temporal profile and polarization}\label{temp}

\begin{figure}
\begin{flushright} \includegraphics[width=10cm]{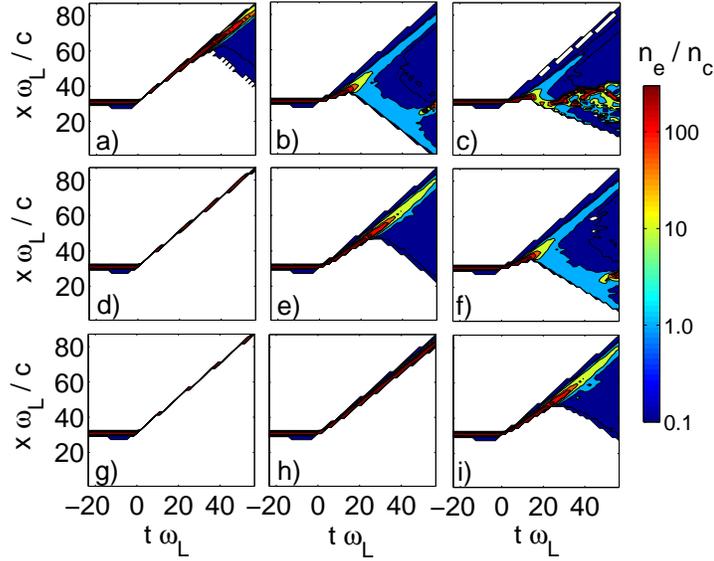} \end{flushright}
\caption{Contour plot of the electron density along the laser propagation axis $x$ and time. The CP laser pulse has field amplitude $a_L=100$ (a,b,c), $a_L=200$ (d,e,f) and $a_L=400$ (g,h,i). Without ramp time (a,d,g), with ramp time $2\,\tau_L$ (b,e,h) and with ramp time $4\,\tau_L$ (c,f,i).}
\label{fig3}
\end{figure}

So far, simulations have been performed using an instantaneous ramp-up of the laser intensity. However, electron response to the laser pulse may be strongly dependent on its temporal profile. We have therefore tested the effect of a finite rise-time of the laser intensity on the electron dynamics and the subsequent light ion acceleration. For this purpose, a target with thicknesses $d_h = 0.08$ (12.7~nm) and $d_l = 0.16$ (25.4~nm) irradiated by varying rise-time laser pulses is considered. Figure~\ref{fig3} shows the temporal evolution of electron density along the laser propagation axis $x$. For times $t < 0$, the laser pulse has not yet reached the target, which undergoes expansion due to non-zero initial temperature only. For times $t > 0$, electrons are strongly accelerated forward by the laser pulse. Once they leave the target, the electrostatic field $a_{tot}$ is built between the electron cloud and ion layers. For sufficiently long times, this field may be strong enough to accelerate backward part or most of the electrons, as previously observed in Fig.~\ref{fig1}b. Considering a laser field amplitude $a_L = 100$ and a flat-top laser intensity profile (Fig.~\ref{fig3}a), a small fraction of the electrons returns toward the ion layers $\sim 5\,\tau_L$ after the beginning of interaction. This time is reduced when the linear ramp-time is increased. For a $2\tau_L$ ramp-time and laser amplitude $a_L = 100$ (Fig.~\ref{fig3}b), a considerable fraction ($> 40\,\%$) of the electrons is accelerated backward less than $4\,\tau_L$ after the beginning of interaction. Increasing the ramp-time up to $4\,\tau_L$ even makes it impossible to extract all electrons from the ion layers, thus preventing efficient light ion acceleration in the linPA. 

Therefore, our proposed mechanism for ion acceleration is very sensitive to the laser pulse profile. It is nonetheless possible to mitigate these effects by increasing the peak laser intensity so as to steepen the intensity temporal profile. Figures~\ref{fig3}d and~\ref{fig3}g indeed show that electrons can be pushed forward as a compressed bunch during the whole simulations by increasing the laser field amplitude up to $a_L = 200$ and $a_L = 400$, respectively. More precisely, for a linear ramp-time $2\,\tau_L$, only a few electrons are accelerated backward when $a_L=200$ (Fig.~\ref{fig3}e), while all electrons are pushed forward for $a_L = 400$ (Fig.~\ref{fig3}h). However, an efficient piston-like acceleration remains difficult to achieve for a $4\,\tau_L$ ramp-time (Fig.~\ref{fig3}f) since it requires a laser field amplitude $a_L = 400$ to prevent electrons from being accelerated backward (Fig.~\ref{fig3}i). 

%\subsection{Influence of the laser polarization}\label{polar}

\begin{figure}
\begin{flushright} \includegraphics[width=10cm]{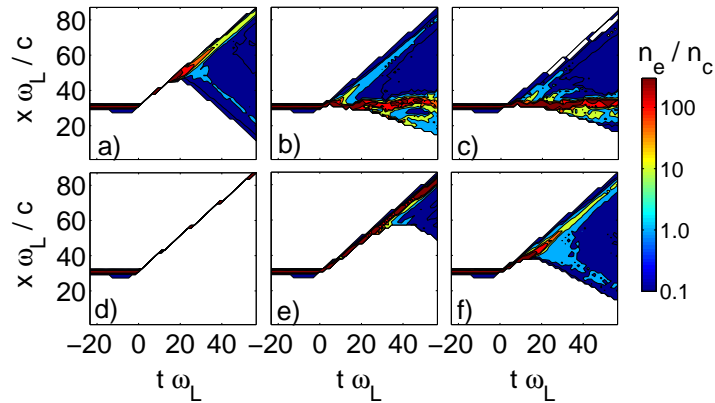} \end{flushright}
\caption{Contour plot of the electron density distribution along the laser propagation axis $x$ and time. The LP laser pulse has field amplitude $a_L=100$ (a,b,c) and $a_L=400$ (d,e,f). Without ramp time (a,d), with ramp time $2\,\tau_L$ (b,e) and with ramp time $4\,\tau_L$ (c,f).}
\label{fig4}
\end{figure}

The influence of the laser polarization has also been investigated. The temporal evolution of the electron density along the laser propagation axis, $x$, when irradiated by a LP laser pulse is presented in Fig.~\ref{fig4} for field amplitudes $a_L = 100$ and $a_L = 400$, and for different ramp-times. When irradiated by LP light, electrons are strongly heated~\cite{macchi_PRL_05,klimo_PRSTAB_08,esirkepov_PRL_04}. To remove all electrons from the target, radiation pressure has to overcome not only electrostatic pressure but also thermal pressure due to extremely hot electrons. Much higher laser intensities are therefore required, which explains why the linPA regime was not relevant under conditions of Ref.~\cite{esirkepov_PRL_02}. At $a_L = 100$ (Figs.~\ref{fig4}a), some electrons are accelerated backward after only $\sim 2.5\,\tau_L$, as compared to $\sim 5\,\tau_L$ when considering a CP laser pulse with similar parameters (Fig.~\ref{fig3}a). Enhancing the linear ramp-time (Figs.~\ref{fig4}b and~\ref{fig4}c) further hinder the electrons' removal by enhancing thermal pressure. This leads us to the conclusion that, for parameters characteristic of this study, and for a given laser intensity, using CP light remains more efficient to extract electrons from the target. 

As can be observed in Figs.~\ref{fig3} and~\ref{fig4}, electrons display a very complex dynamical behavior. In particular, the early stage of the laser pulse interaction with the target is characterized by a strong acceleration of electrons, which gain relativistic energies in less than one optical cycle. Under such conditions, the Doppler-shift is modified and therefore the radiation pressure. Furthermore, due to the large field amplitude ($a_L>100$ typically) and the small target thickness, the electron cloud is not opaque to the laser. Relativistically induced transparency occurs~\cite{relat_transp}, and the laser pulse penetrates through the whole electron cloud, even though this cloud expands on several wavelengths. Obviously, this also modifies the radiation pressure. While these phenomena are accounted for in our PIC simulations, radiation friction is not taken into account. It may also strongly influence the electron dynamics and in turn ion acceleration, as previously observed in Ref.~\cite{naumova_PRL_08}. Modeling the electron dynamics under such conditions thus appears to be very challenging and we leave this complex issue for future investigations.

To conclude with the electrons' behavior, it must be stressed that both 1D and 2D simulations overestimate the electrostatic field between the electron and ion layers, thus causing the electrons to remain bound to the ion layers. Conversely, in 3D geometry, electrons are able to separate definitely from the ion layers when their kinetic energy $U_k = \gamma_e-1$ becomes larger then their potential energy in the ions field, $U_p = Q/(4\,\pi\,d)$, where $Q \sim a_h\,w_{\perp}^2$ is the total charge of the electron cloud in units of $e\,n_c/k_L^3$ and $d$ is the distance between the electron cloud and the heavy ion layer~\cite{tikhonchuk_POP_02}. 

From previous simulations, one can expect electrons to reach kinetic energies $10 < U_k \le 100$ ($\sim 5 - 50$~MeV) after only a few optical cycles in the laser field (e.g. see Fig.~\ref{fig1}c). Considering that electrons can be efficiently pushed over distances of a few wavelengths on such time scales, generation of electron bunches with the charge a few nC is expected under current conditions of irradiation and target thickness.

%%%%%%%%%%%%%%%%%%%%%%%%%%%%%%%%%%
% MULTI-D EFFECTS & SCALING LAWS %
%%%%%%%%%%%%%%%%%%%%%%%%%%%%%%%%%%
\section{Multi-dimensional effects on the linPA and directed Coulomb explosion (DCE) of the target}\label{sec3}

Previous modeling and numerical simulations rely on an idealized 1D picture of the linPA. In this Section, 2D simulations are discussed that show how multi-dimensional effects arise when electrons are pushed far enough from the ion layers. Their impacts on ion acceleration are discussed. Especially, we show that linPA is only the first stage of the acceleration process. A second acceleration stage occurs when the distance between electrons and ions becomes similar to the transverse size of the system.

\subsection{Two-dimensional numerical simulations}\label{simu2d}

Two-dimensional simulations have been performed with CALDER to investigate multi-dimensional effects on the proposed acceleration scheme. In these simulations, the CP laser pulse has a flat-top temporal profile with duration $\tau_p = 10\,\tau_L$ ($\sim 33$~fs), and the maximum field amplitude is $a_L=100$ ($\sim 1.4 \times 10^{22}\,{\rm W/cm^2}$). Along the transverse $y$-direction, the laser intensity follows a 6th-order super-Gaussian distribution. Two value of the laser focal spot full-width at half-maximum have been considered: $w_L=10\,\lambda_L$ and $w_L=20\,\lambda_L$ ($\sim 10 - 20\,{\rm \mu m}$, respectively). This laser pulse is focused on an ultra-thin double layer target following the design considerations of the previous sections. The target's first layer is made of carbon with atomic density $n_h = 58$ ($\sim 6.4 \times 10^{22}\,{\rm cm^{-3}}$), thickness $d_h = 0.08$ ($\sim 12.7$~nm) and transverse width $w_h = 200$ ($\sim 30\,{\rm \mu m}$). An hydrogen dot with thickness $d_l = 0.16$ ($\sim 25.4$~nm) and transverse width $w_l = \lambda_L$ ($\sim 1\,{\mu m}$) is placed on the rear-side of the first layer. Two different atomic densities, $n_l = 6$ and $n_l = 12$ ($\sim 0.6 - 1.2 \times 10^{22}\,{\rm cm^{-3}}$, respectively), have been considered.

\begin{figure}
\begin{flushright} \includegraphics[width=14cm]{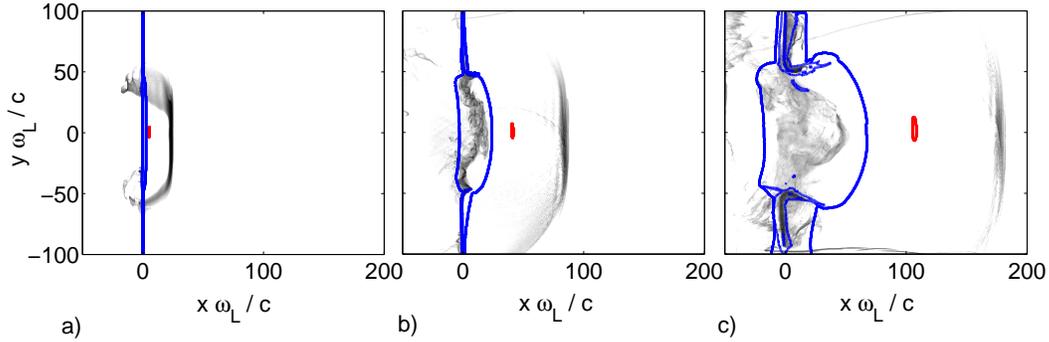} \end{flushright}
\caption{Snapshots of the electron distribution (gray scale), carbon contour plot (blue) and proton contour plot (red) for different times: a) $t=4\,\tau_L$ after the beginning of the interaction; b) $t=14\,\tau_L$ after the beginning of the interaction and c) $t=30\,\tau_L$ after the beginning of the interaction.}
\label{fig5}
\end{figure}

\begin{figure}
\begin{flushright} \includegraphics[width=11cm]{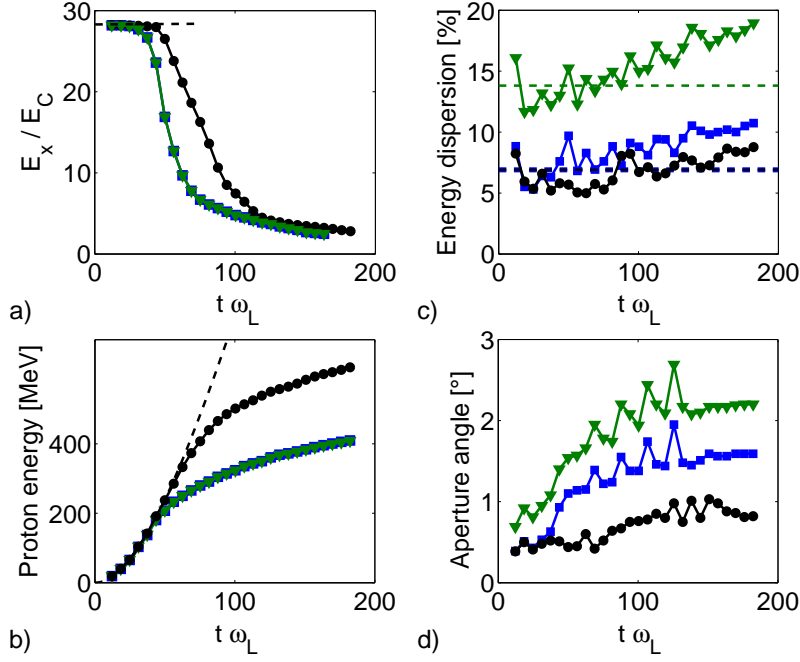} \end{flushright}
\caption{Temporal evolution of: a) the accelerating electrostatic field along the $x$-direction at the position of the proton bunch (the dashed curve accounts for the electrostatic field $a_h$); b)~the proton energy; c)~the energy dispersion; d)~the proton beam angular aperture. For $n_l=6$, $w_L=10\,\lambda_L$ (blue, square), $n_l=12$, $w_L=10\,\lambda_L$ (green, triangle) and $n_l=6$, $w_L=20\,\lambda_L$ (black, circle). In panel b)~and~c), dashed curves account for predictions from the linPA model.}
\label{fig6}
\end{figure}

The simulations results are given in Figs.~\ref{fig5} and~\ref{fig6}. Figure~\ref{fig5} shows the spatial distributions of electrons, carbon ions and protons at three different times with the parameters $n_l = 6$ and $w_L = 10\,\lambda_L$. In Fig.~\ref{fig5}a, the characteristic features of the linPA are clearly apparent. Four laser periods after the beginning of the interaction, an electron bunch with the transverse width $w_{\perp} \sim 12\,\lambda_L$, of the order of the laser focal spot, is separated from the ion layers. It is pushed forward over a distance of the order of $4\,\lambda_L$ (consistent with the previous consideration that electrons leave the target with the light velocity). A strong electrostatic field, $a_h \sim 28$ ($\sim 90\,{\rm TV/m}$), is built up between the electron and carbon layers. This is confirmed in Fig.~\ref{fig6}a, where the longitudinal component of the electrostatic field at the position of the proton bunch is shown as a function of time. The early stage of proton acceleration is characterized by a constant accelerating field, and its duration, $t_{linPA}$, can be extracted from Fig.~\ref{fig6}a. For a laser pulse transverse width $w_L = 10\,\lambda_L$, one has $t_{linPA} \sim 7\,\lambda_L$, while $t_{linPA} \sim 9.5\,\lambda_L$ for $w_L = 20\,\lambda_L$. At this point, it is interesting to note that $t_{linPA}$ is not twice as long for $w_L = 20\,\lambda_L$ as for $w_L = 10\,\lambda_L$. This is because in the case where $w_L = 20\,\lambda_L$, some electrons come back toward the ion layers and thus reduce the accelerating field prematurely. For short times, $t \lesssim t_{linPA}$, protons are consequently accelerated in the capacitor-like electrostatic field characteristic of the linPA process. Figure~\ref{fig6}b shows the temporal evolution of the proton energy for different target parameters. As predicted from the linPA model, the proton energy during this stage depends neither on the second layer's density $n_l$, nor on the first layer's width. In this early stage, an excellent agreement with theoretical predictions is obtained. 

In addition, Figs.~\ref{fig6}c and~\ref{fig6}d show the temporal evolution of the proton bunch energy and angular dispersions. Especially, Fig.~\ref{fig6}c shows that, while the proton energy dispersion is not strongly modified by increasing the laser pulse transverse width, it can be tuned by changing the density of the hydrogen dot. This confirms that the energy dispersion follows from electrostatic repulsion inside the proton bunch itself. Energy dispersions of the order of $7\,\%$ and $14\,\%$ are obtained depending on the density $n_l$, in very good agreement with predictions from the linPA model. Similar observations follow from Fig.~\ref{fig6}d concerning the angular dispersion of the proton bunch. In the linPA stage of proton acceleration, the angular aperture of the proton beam is therefore mainly governed by electrostatic self-repulsion. For parameters of this study, angular aperture of a few degree only are obtained. 

Let us now investigate proton acceleration on times larger than $t_{linPA}$. As can be observed in Figs.~\ref{fig5}b and~\ref{fig5}c, electrons are here pushed over a distance of the order of, or larger than $w_{\perp}$. Multi-dimensional effects thus set in, reducing the accelerating field (Fig.~\ref{fig6}a) and making proton acceleration less efficient (Fig.~\ref{fig6}b). While the proton energy in the linPA stage of ion acceleration did not depend on the transverse width $w_L$ of the laser pulse, increasing $w_L$ allows to enhance the final proton energy by delaying multi-dimensional effects. Furthermore, in this second stage of proton acceleration, one can observe an enhancement of both the energy and angular dispersion of the proton bunch (Figs.~\ref{fig6}c and~\ref{fig6}d). While energy dispersion is still mainly determined by the charge in the second layer (Fig.~\ref{fig6}c), one can observe that the proton beam angular aperture depends on both the carbon layer's width and the density of the hydrogen dot (Fig.~\ref{fig6}d). This prompts us to suggest that, in this phase of the acceleration process, both the transverse inhomogeneity of the accelerating field and Coulomb self-repulsion are responsible for the proton bunch angular aperture. At the end of the simulation, energy dispersion of the order of $10\,\%$ and angular aperture $\lesssim 3^{\circ}$ are obtained. 

These simulations demonstrate that even when multi-dimensional effects are accounted for, the linPA mechanism is effective for accelerating light ion beams, as the first, transient, stage in a two-step acceleration process. In 3D geometry, the duration of the linPA stage might be further reduced thus limiting the final proton energy. On the other hand, one expects 3D effects to mitigate the energy and angular dispersions.

\subsection{Estimates for three-dimensional effects}

To understand the behavior of light ions over times larger than $t_{linPA}$, it is necessary to compute the distance traveled during the linPA stage. The position of the slowest and fastest light ions at $t=t_{linPA}$ can be easily obtained by integrating the relativistic equations of motion of the light ions in the accelerating fields $a_h$ and $a_{tot}$, respectively. From this we obtain the traveled distance $x_l$ and the light ion layer thickness $\delta x_l$ at the end of the linPA stage:
\begin{eqnarray}
\label{eq5} x_l \sim \sqrt{t_r^2+t_{linPA}^2} - t_r\,,\\
\label{eq6} \delta x_l \sim d_l + \sqrt{t_r^2/\big(1+a_l/a_h\big)^2+t_{linPA}^2}-\sqrt{t_r^2+t_{linPA}^2}-(a_l/a_h)\,t_r\,.\,\,\,
\end{eqnarray}
In what follows, we shall show that light ions exhibit two distinct long-time behaviors depending on whether or not they are relativistic at the end of the linPA stage.

(i) As discussed in Sec.~\ref{sec2}, for effective laser powers below $P_L^{(r)}$ (which also corresponds to $t_{linPA} \ll t_r$), light ions are non-relativistic at the end of the linPA stage. From Eqs.~(\ref{eq5}) and~(\ref{eq6}), we obtain that such ions have propagated on a distance $x_l^{(c)} \sim t_{linPA}^2/(2\,t_r)$ during this first stage, while the light ion layer thickness has increased to $\delta x_l^{(c)} \sim d_l + (a_l/a_h)\,t_{linPA}^2/(2\,t_r)$. For most parameters of interest, this thickness $\delta x_l^{(c)}$ remains much smaller than the transverse width $w_l$, so that the light ion layer conserves its pancake-like shape. Moreover, because $x_l^{(c)} \ll t_{linPA} \lesssim w_{\perp}$, light ions remains close to the heavy ion layer at the end of the non-relativistic linPA stage. As a consequence, they will gain further energy in the electrostatic field due to heavy ions while the effect of the electron cloud on ion acceleration for $t > t_{linPA}$ can be neglected. The acceleration process then becomes similar to DCE, with the difference that particles have a finite initial kinetic energy. Considering that both light and heavy ion layers retain their disk-like shape and that the distance between both layers $\delta x_l^{(c)} \ll t_{linPA}$ remains small compared to the transverse width $w_{\perp}$, the accelerating field at the rear side of the heavy ion layer behaves as $a_{DCE}(x) = (a_h/2)\,(1-2\,x/w_{\perp})$. In Ref.~\cite{esirkepov_PRL_02}, the authors assume that the field inhomogeneity $\Delta a_h \sim -a_h\,d_l^{(c)}/w_{\perp}$ through the second layer is responsible for both, light ions bunching and energy dispersion. However, they do not account for the self-consistent field $a_l$ in the second layer that, under conditions of interest in our work, exceeds the field inhomogeneity $\Delta a_h$. We thus find that the field inhomogeneity weakly affects the long-time ($t \gg t_{linPA}$) energy dispersion of light ions, which mainly originates from electrostatic repulsion between the protons, as previously observed in 2D simulations (Sec.~\ref{simu2d}). 

The energy gain of light ions during the DCE stage can be estimated from their potential energy in the heavy ion electrostatic field, while their energy dispersion is obtained from their potential energy in the self-consistent field $a_l$:
\begin{eqnarray}
\label{eq7} \Epsilon_{DCE}       \sim Z_l\,a_h\,w_{\perp}/4 - \Epsilon_{linPA}/2\,,\\
\label{eq8} \Delta\Epsilon_{DCE} \sim Z_l\,a_l\,w_l/4\,.
\end{eqnarray}
The first term in the right-hand-side of Eq.~(\ref{eq7}) is similar to estimates obtained in Ref.~\cite{esirkepov_PRL_02}. The second term accounts for the fact that, in contrast to what occurs in ``classical'' DCE, part of the potential energy of light ions has already been transformed into kinetic energy in the linPA stage. Obviously, because light ions have not traveled far from the heavy ions layer during the non-relativistic linPA stage under current conditions, the DCE stage provides the main contribution to the final energy of accelerated ions. Finally, the light ion energy at $t \gg t_{linPA}$ is simply $\sim Z_l\,a_h\,w_{\perp}/4$ and it scales as the square-root of the laser power. Moreover, the relative energy dispersion $\sim w_l\,a_l/(w_{\perp}\,a_h)$ is obtained from Eqs.~(\ref{eq7}) and~(\ref{eq8}). It can be kept to a rather low level assuming that both the electrostatic field $a_l$ in the light ion layer and the transverse width $w_l$ of this layer are small compared to the accelerating field $a_h$ and the transverse width $w_{\perp}$, respectively.

In addition, we want to point out that previous considerations on 3D electron behavior on long times (Sec.~\ref{temp}) suggest that the ion beam leaves the target as a non-neutral bunch. The effect of electrons on long time scales, which can be disastrous in terms of energy dispersion, can thus be neglected: Coulomb repulsion of the light ions is the main source of energy dispersion. 

(ii) For laser powers above $P_L^{(r)}$, light ions gain relativistic energies in the linPA stage and the acceleration process beyond $t_{linPA}$ is modified. Indeed, taking $t_{linPA} \gg t_r$ in Eqs.~(\ref{eq5}) and~(\ref{eq6}), we obtains that light ions have traveled over a distance $x_l^{(ur)} \sim t_{linPA} - t_r$ in the linPA, while their thickness as increased to $\delta x_l^{(ur)} \sim d_l + (a_l/a_h)\,t_r$. Once more, for characteristic parameters of interest, $\delta x_l^{(ur)} \ll w_l$, and the light ion bunch keeps its pancake-shape. However, considering that electrons have propagated on a distance $x_e \sim t_{linPA}$, light ions should stay close to the electron cloud at the end of the linPA stage ($x_e - x_l^{(ur)} \sim t_r \ll w_{\perp}$). Therefore, in a second stage, light ions may gain further energy in the field due to the electron cloud, which is continuously pushed by the laser pulse. 

This scenario might prove very interesting for generating high-quality relativistic ion beams. Nevertheless, a full 3D modeling is required to quantitatively account for the non-trivial long-time electron behavior. This will be the subject of future studies.

\section{Discussion and conclusion}\label{sec5}

In this paper, we propose a novel ion acceleration mechanism allowing for an excellent control of all ion beam properties. This mechanism is based on complete removal of electrons from an ultra-thin double layer target by radiation pressure. The consequences of this mechanism in terms of ion acceleration are discussed here for the first time. Numerical simulations show that complete expulsion of electrons from the target can be achieved above a laser intensity threshold that depends on the areal charge of the irradiated target as well as on the pulse profile and polarization. For a nano-scale target irradiated by a CP laser pulse, laser field amplitudes $a_L > 50$, typically, are required. 

We discuss acceleration of light ions of a double-layer target, in a regime where a capacitor-like, quasi-homogeneous, electrostatic field is built up between the ion and electron layers. The resulting acceleration process is referred to as linPA. It may accelerate light ions up to high energies over a few optical cycles only, with an energy dispersion that can be controlled by the target properties (areal charge and geometry). Generation of relativistic light ions during this stage is also expected when using an effective laser power larger than $P_L^{(r)}$, proportional to the squared inverse charge-over-mass ratio. For protons, this threshold power is of a few petawatt. 

Beyond this transient linPA stage, the light ion acceleration is determined by multi-dimensional effects. Analytical considerations suggest that two scenarii should be considered depending on whether or not light ions have acquired relativistic velocities at the end of the linPA stage. While low energy ion generation mainly follows from DCE, relativistic ions are mainly accelerated in the linPA stage and may gain further energy in the electron cloud field afterwards. In between, modeling from the linPA should provide good estimates for the ion source properties. The energy dispersion of the resulting ion source follows mainly from Coulomb self-repulsion of the light ions. Energy dispersion is shown to depend mainly on the ratio of the areal charge in the two layers and on the target geometry. An excellent control of the energy, energy dispersion as well as total charge of the light ion beam can thus be achieved by choosing appropriate target properties. This approach thus appears to be extremely attractive for applications such as proton therapy, where high-quality ion beams are required.  

Multi-dimensional effects are partly accounted for in our theoretical model, but, at the present stage, we can only give rough estimates on expected ion beam properties. Moreover, we do not discuss the complex behavior of electrons when they are strongly accelerated by the laser pulse. In a realistic 3D configuration, inhomogeneities in the laser intensity distribution will definitely affect this process. As already observed by other authors~\cite{klimo_PRSTAB_08,robinson_qiao,pegoraro_PRL_07}, Rayleigh-Taylor-like or Weibel instabilities may occur. Another mechanism likely to influence the electron behavior and in turn ion acceleration is the radiation friction experienced by ultra-relativistic electrons in the presence of a high-amplitude laser field. All these effects on ion acceleration are beyond the scope of this work and are left for future investigations. 

The drawbacks of the proposed light ion acceleration scheme are similar to those of other methods based on the irradiation of a thin double-layer target by high power lasers. First, maintaining the integrity of such thin targets requires ultra-high contrast laser pulses (in excess of $10^{11}$). Also, the conversion efficiency from the laser to light ion is rather small and most of the absorbed laser energy goes to electrons and heavy ions. Moreover, techniques to separate the high-quality light ion beam for other particles such as heavy ions and electrons must be considered. While this is greatly simplified by the fact that all particles have different charge-over-mass ratio, standard deflection and shielding techniques may drastically enhance the size of the ion source. 

Experimental exploration of this acceleration mechanism requires petawatt lasers with well-controlled temporal profile and the design of nano-structured targets, such as diamond-like carbon foils. Fast developments in the technology of high-power lasers~\cite{laser-dev} and ultra-thin target fabrication~\cite{mccomas_RSI_04} may make generation of high-quality, well-controlled, light ion beams feasible in a near future.  Finally, let us note that the possibility to obtain high-energy ions on very short time scales (a few optical cycles) makes our mechanism particularly interesting for investigation at high-power, modest energy, laser facilities such as ELI~\cite{ELI}.

%%%%%%%%%%%%%%%%%%%%%
%    CONCLUSIONS    %
%%%%%%%%%%%%%%%%%%%%%
\section*{Acknowledgments}

The authors are grateful to V.~T.~Tikhonchuk, A.~Macchi and T.~Cowan for fruitfull discussions and pertinent comments.

%%%%%%%%%%%%%%%%
% BIBLIOGRAPHY %
%%%%%%%%%%%%%%%%
%\newpage $\,$


\section*{References}

\begin{thebibliography}{2}

\bibitem{borghesi_POP_02} Borghesi M {\it et al} (2002) Phys. Plasmas {\bf 9}, 2214. 

\bibitem{roth_PRL_01} Roth M {\it et al.} (2001) Phys. Rev. Lett. {\bf 86}, 436.

\bibitem{temporal_POP_02} Temporal M, Honrubia J J and Atzeni S (2002) Phys. Plasmas {\bf 9}, 3098.

\bibitem{naumova_PRL_08} Naumova N, Schlegel T, Tikhonchuk VT, Labaune C, Sokolov I V and Mourou G (2009) Phys. Rev. Lett. {\bf 102}, 025002.

\bibitem{koshorovEJP_bulanovPPR} Khoroshkov V S and Minakova E I (1998) Eur. J. Phys. {\bf 19}, 523; Bulanov S V and Khoroshkov V S (2002) Plasma Phys. Rep. {\bf 28}, 453 (2002).

\bibitem{snavely00_robson06} Snavely R A {\it et al.} (2000) Phys. Rev. Lett. {\bf 85}, 2945; Robson L {\it et al.} (2006) Nature Physics {\bf 3}, 58.

\bibitem{sentoku_POP_03} Sentoku Y, Cowan T E, Kemp A, Ruhl H (2003) Phys. Plasmas {\bf 10}, 2009.

\bibitem{shock_acceleration} Denavit J (1992) Phys. Rev. Lett. {\bf 69}, 3052; Silva L, Marti M, Davies J R, Fonseca R A, Ren C, Tsung F S and Mori W B (2004) Phys. Rev. Lett. {\bf 92}, 015002.

\bibitem{macchi_PRL_05} Macchi A, Cattani F, Liseykina T V and Cornolti F (2005) Phys. Rev. Lett. {\bf 94}, 165003.

\bibitem{solitary_wave} Zhidkov A, Uesaka M, Sasaki A and Daido H (2002) Phys. Rev. Lett. {\bf 89}, 215002.

\bibitem{wilks_POP_01} Wilks S C {\it et al.} (2001) Phys. Plasmas {\bf 8}, 542.

\bibitem{fuchs_PRL_05} Fuchs J {\it et al.} (2005) Phys. Rev. Lett. {\bf 94}, 045004.

\bibitem{zhang_POP_07} Zhang X, Shen B, Li X, Jin Z and Wang F (2007) Phys. Plasmas {\bf 14}, 073101.

\bibitem{klimo_PRSTAB_08} Klimo O, Psikal J, Limpouch J and Tikhonchuk V T (2008) Phys. Rev. ST Accel. Beams {\bf 11}, 031301.

\bibitem{robinson_qiao} Robinson A P L, Zepf M, Kar S , Evan R G and Bellei C (2008) New J. Phys. {\bf 10}, 013021; Qiao B, Zepf M and Borghesi M (2009) Phys. Rev. Lett. {\bf 102}, 145002. 

\bibitem{yan_rykovanov_eliasson} Yan X Q, Lin C, Sheng Z M, Guo Z Y, Liu B C, Lu Y R, Fang J X and Chen J E (2008) Phys. Rev. Lett. {\bf 100}, 135003; Rykovanov S G, Schreiber J, Meyer-ter-Vehn J, Bellei C, Henig A, Wu H C and Geissler M (2008) New. J. Phys. {\bf 10}, 113005; Eliasson B, Chuan S L, Shao X, Sagdeev R Z and Shukla P K (2009) New J. Phys. {\bf 11}, 073006. 

\bibitem{macchi_PRL_09} Macchi A, Veghini S and Pegoraro F, {\it ``Light Sail'' Acceleration Revisited}, accepted for publication in Phys. Rev. Lett..

\bibitem{esirkepov_PRL_02} Esirkepov T Zh {\it et al.} (2002) Phys. Rev. Lett. {\bf 89}, 175003.

\bibitem{schwoerer_NATURE_06} Schwoerer H, Pfotenhauer S, J\"{a}ckel O,  Amthor K U, Liesfeld B, Ziegler W, Sauerbray R, Ledingham K W D and Esirkepov T Zh (2006) Nature {\bf 439}, 445.

\bibitem{hegelich_NATURE_06} Hegelich B M, Albright B J, Cobble J, Flippo K, Letzring S, Paffett M., Ruhl H, Schreiber J, Schulze R K and Fern\'{a}ndez J C (2006) Nature {\bf 439}, 441.

\bibitem{albright_PRL_06} Albright B J, Yin L, Hegelich B M, Bowers K J, Kwan T J T and Fern\'{a}ndez (2006) Phys. Rev. Lett. {\bf 97}, 115002.

\bibitem{teravetisyan_PRL_06} Ter-Avetisyan S, Schn\"{u}rer M, Nickles P V, Kalashnikov M, Risse E, Sokollik T, Sandner W, Andreev A and Tikhonchuk V T T (2006) Phys. Rev. Lett. {\bf 96}, 145006.

\bibitem{brantov_POP_06} Brantov A V, Tikhonchuk V T, Klimo O, Romanov D V, Ter-Avetisyan S, Schn\"{u}rer M, Sokollic T and Nickles P V (2006) Phys. Plasmas {\bf 13}, 122705.

\bibitem{DCE} Fourkal E, Velchev I and Ma C M (2005) Phys. Rev. E {\bf 71}, 036412; Bulanov S S {\it et al.} (2008) Phys. Rev. E~{\bf 78}, 026412.

\bibitem{lefebvre_NF_03} Lefebvre E {\it et al} (2003) Nucl. Fusion {\bf 43}, 629.

\bibitem{rykovanov_NJP_08} Rykovanov S G, Schreiber J, Meyer-ter-Vehn J, Bellei C, Henig A, Wu H C and Geissler M (2008) New. J. Phys. {\bf 10}, 113005.

\bibitem{esirkepov_PRL_04} Esirkepov T Zh, Borghesi M, Bulanov S V, Mourou G and Tajima T (2004) Phys. Rev. Lett. {\bf 92}, 175003.

\bibitem{relat_transp} Lefebvre E and Bonnaud G (1995) Phys. Rev. Lett. {\bf 74}, 2002; Cattani F,  Kim A, Anderson D and Lisak M (2000) Phys. Rev. E {\bf 62}, 1234.

\bibitem{tikhonchuk_POP_02} Tikhonchuk V T (2002) Phys. Plasmas {\bf 9}, 1416.

\bibitem{pegoraro_PRL_07} Pegoraro F and Bulanov S V (2007) Phys. Rev. Lett. {\bf 99}, 065002.

\bibitem{kulagin} Kulagin V V, Cherepenin V A and Suk H (2004) Phys. Plasmas {\bf 11}, 5239; Kulagin V V, Cherepenin V A, Hur M S and Suk H (2007), Phys. Rev. Lett. {\bf 99}, 124801.

\bibitem{steiger_PRA_72} Steiger A D and Woods C H (1972) Phys. Rev. A {\bf 5}, 1467; Zhidkov A, Koga J, Sasaki A and Uesaka M (2002) Phys. Rev. Lett. {\bf 88}, 185002. 

\bibitem{laser-dev} Shah R C, Johnson R P, Shimada T, Flippo K A, Fern\'{a}ndez J C and Hegelich B M (2009) {\it High-temporal contrast using low-gain optical parametric amplification}, Opt. Lett, accepted.

\bibitem{mccomas_RSI_04} McComas D J, Allegrini F, Pollock C J, Funsten H O, Ritzau S and Goeckler G (2004) Rev. Sci. Instrum. {\bf 75}, 4863.

\bibitem{ELI} http://www.extreme-light-infrastructure.eu.

\end{thebibliography}
\end{document}